\documentclass[fleqn,10pt]{wlscirep}
\usepackage[utf8]{inputenc}
\usepackage[T1]{fontenc}
\usepackage{makecell}
\usepackage{nicematrix}
\usepackage{enumitem}
\usepackage{amsthm}
\usepackage{amssymb}
\usepackage{bm}
\usepackage{cite}
\usepackage{float}
\usepackage{mathrsfs}
\usepackage{amsfonts}
\usepackage{bm}
\usepackage{float}
\usepackage{mathrsfs}
\usepackage{graphicx}
\usepackage{algorithm, algpseudocode}
\usepackage{subfigure}
\usepackage{tabularx}
\usepackage{flushend}
\usepackage{epsfig}
\usepackage{cite}
\usepackage[super]{nth}
\usepackage{flushend}
\usepackage{multirow}
\usepackage{color}
\usepackage{enumerate}
\usepackage{easylist}
\usepackage{kantlipsum}
\usepackage{amsmath}
\usepackage[utf8]{inputenc}

\title{HistoSegCap: Capsules for Weakly-Supervised Semantic Segmentation of Histological Tissue Type in Whole Slide Images}

\author[1,+]{Mobina Mansoori}
\author[1,+]{Sajjad Shahabodini}
\author[2]{Jamshid Abouei}
\author[1,*]{Arash Mohammadi}
\author[3]{Konstantinos N. Plataniotis}

\affil[1]{Concordia Institute for Information Systems Engineering, Concordia University, Montreal, QC, Canada}
\affil[2]{Department of Electrical Engineering, University of Yazd, Yazd, Iran}
\affil[3]{Department of Electrical and Computer Engineering, University of Toronto, Toronto, Canada}
\affil[*]{arash.mohammadi@concordia.ca}

\affil[+]{these authors contributed equally to this work}

\begin{abstract}
{Digital pathology involves converting physical tissue slides into high-resolution Whole Slide Images (WSIs), which pathologists analyze for disease-affected tissues. However, large histology slides with numerous microscopic fields pose challenges for visual search. To aid pathologists, Computer Aided Diagnosis (CAD) systems offer visual assistance in efficiently examining WSIs and identifying diagnostically relevant regions. This paper presents a novel histopathological image analysis method employing Weakly Supervised Semantic Segmentation (WSSS) based on Capsule Networks, the first such application. The proposed model is evaluated using the Atlas of Digital Pathology (ADP) dataset and its performance is compared with other histopathological semantic segmentation methodologies. The findings underscore the potential of Capsule Networks in enhancing the precision and efficiency of histopathological image analysis. Experimental results show that the proposed model outperforms traditional methods in terms of accuracy and the mean Intersection-over-Union (mIoU) metric.}
\end{abstract}

\begin{document}
\flushbottom
\maketitle
\thispagestyle{empty}

\section*{Introduction}
\indent Histopathology with Whole Slide Imaging (WSI) is considered the gold standard method for producing high-resolution images from glass slides \cite{faherty2023role}. In the realm of clinical practice, small sections of tissue are stained with Hematoxylin and Eosin and subsequently examined under a microscope by pathologists, who rely on their expertise  and experience to evaluate cell and tissue characteristics, including morphology and cytology \cite{lu2020prognostic,sharma2023hyaline}. Pathologists then sift through glass slides to identify abnormal regions referred to as Regions of Interest (ROI). These ROIs play a crucial role in clinical prediction such as prognosis, diagnosis, and metastasis \cite{zhou2019cgc,yu2016predicting,ahmedt2022survey}. Typically, pathologists meticulously analyze numerous glass slides each day, and the accuracy of diagnosis is contingent upon individual factors such as pathologists' fatigue and level of experience. However, challenges persist within this domain. For example, when examining breast biopsies, pathologists have exhibited a 24.7\% discrepancy rate in their diagnoses \cite{elmore2015diagnostic}. In response to such challenges, Computer Aided Diagnosis (CAD) systems have been introduced to assist pathologists in refining their diagnoses, enhancing accuracy by scanning through glass slides, and speeding up the diagnostic process. CAD leverages computer-generated output as an auxiliary tool for clinicians, aiding them in making precise diagnoses. This approach differs from fully automated computer diagnosis, which relies solely on computer algorithms. Nevertheless, the development of effective and adaptable computational pathology tools is impeded by limited access to histological images annotated at the pixel level and the need for advanced computer vision training using supervised learning methods \cite{li2022comprehensive}.

Digital pathology slide images are significantly large and necessitate patch-level segmentation. Furthermore, the majority of databases comprise annotated and semantically segmented histopathological images at the patch level. Thus, in this study, patch-level annotation of Histological Tissue Type (HTT) is deemed more appropriate, and a semantic segmentation algorithm is developed to conduct patch-level annotation on various bodily organs, thereby enabling more precise and specific diagnoses \cite{bokhorst2023deep,yacob2023weakly,chan2019histosegnet}. 

Patch-level annotation models belong to the category of Weakly Supervised Semantic Segmentation (WSSS) and rely on global labels. Unlike fully supervised semantic segmentation, WSSS is a cost-effective and time-efficient approach \cite{lin2023clip}. Deep models are trained using WSIs from slides for weakly-supervised methods, enabling satisfactory pixel segmentation predictions without the requirement of detailed local annotations through multi-instance learning. In WSSS, a small portion of labeled data is utilized for learning, while a substantial amount of unlabeled data is employed to enhance the learning rate and achieve the final outcome \cite{yan2022deep}. 

Recent studies have employed Convolutional Neural Networks (CNNs) for the semantic segmentation of HTT \cite{chan2019histosegnet}. However, CNN-based learning and training approaches come with several drawbacks. Firstly, CNNs operate without pre-processing and require no prior knowledge of features or types. Although CNNs possess substantial learning capabilities, they are less effective when dealing with large datasets \cite{afshar2018brain}. Moreover, CNNs do not provide perfect results as they are not resistant to dependent transformation and do not account for spatial relationships within the images. The max-pooling operation in CNNs further leads to a loss of fine spatial information and the potential discarding of relevant information \cite{bonheur2019matwo}. To address these limitations, \cite{sabour2017dynamic} introduced Capsule Networks, also known as dynamic routing. Capsule Networks consist of multiple neurons whose activity vectors determine the position and orientation of Capsules. The length of the vector represents the probability that a specific object is represented by the network. The most important characteristic of dynamic routing is routing by agreement, where lower-level Capsules predict the outcome of higher-level Capsules, activating the latter only if these predictions are correct. It is worth mentioning that dynamic routing exhibits high sensitivity to image backgrounds, resulting in improved accuracy in classifying segmented tissues \cite{afshar2019capsule}. To the best of the authors' knowledge, this study represents the first application of Capsules for Weakly Supervised Semantic Segmentation and provides an interpretation of dynamic routing.

In order to accurately analyze histopathological images, it is crucial to assign a semantic label to each pixel of a WSI. Therefore, histopathology images can be visually classified according to tissue types. To achieve this, WSIs must be categorized into tissues based on morphological and functional modes, as well as non-tissues such as Background and Other. This accurate labeling of histopathological images enables the development of computer-aided diagnosis systems, which can assist pathologists in detecting and diagnosing various diseases with higher precision and efficiency. So far, various approaches have been proposed to implement WSSS models using CNNs. Nevertheless, existing literature lacks any model that employs Capsule Networks for WSSS techniques. To address this gap, this paper introduces a Capsule-based architecture aimed at surpassing the limitations of CNN models and enhancing WSSS performance. The main contributions of the paper are summarized as~follows:
\begin{itemize}
\item	In this study, we propose an innovative model utilizing Capsule Networks for the semantic segmentation of histopathological images, accompanied by the first-ever interpretation method specifically designed for Capsule Networks.
\item	As a new method to improve semantic segmentation results, we leverage the reconstruction layers of the Capsule Networks to identify different labels’ locations within the input images.
\item	The effectiveness of the model is assessed on the Atlas of Digital Pathology (ADP) dataset \cite{hosseini2019atlas}, comprising histopathological images from different organs.
\item   A comparative analysis is conducted between the performance of the proposed HistoSegCap model and alternative methods for histopathological semantic segmentation, showcasing its superior performance. Moreover, we demonstrate the superiority of Capsule models over CNNs in terms of accuracy and WSSS results.
\end{itemize}

\section*{Related Work}
\subsection*{Dynamic Routing} Capsule Networks represent a relatively new type of neural network architecture that exhibits promising potential in various computer vision tasks, including image segmentation. Unlike traditional CNNs, Capsule Networks incorporate dynamic routing to effectively handle spatial relationships between features within an image \cite{sabour2017dynamic,hinton2011transforming}. By utilizing Capsules, these networks excel at preserving spatial information and addressing occlusion-related challenges, making them particularly advantageous for image segmentation tasks. Notably, in recent years, numerous studies have investigated the application of Capsule Networks in image segmentation, leading to promising results. For instance, researchers have employed Capsule Networks to segment medical images like MRI scans for brain tumor segmentation \cite{wang2018weakly,xu2016deep} and pathological lung segmentation \cite{turkki2016antibody}. To the best of our knowledge, there is currently no existing report that has explored the utilization of Capsule Networks in conjunction with WSSS.

\subsection*{Weakly-Supervised Semantic Segmentation} 
Fully supervised learning is widely recognized as a highly accurate approach for semantic segmentation since it trains at the pixel level. However, this method necessitates annotations, which can be time-consuming and costly. To tackle this challenge, weakly supervised learning has emerged as an alternative to fully supervised learning. Weakly supervised learning techniques can be categorized into four groups: (a) Multi-Instance Learning (MIL) methods, which optimize the process of assigning a pixel to each image label for every image. Previous studies have demonstrated the effectiveness of MIL methods in segmentation tasks \cite{hashemzehi2020detection,nogueira2021deep}. (b) Graphical model-oriented methods, which detect uniform-looking regions and predict latent variable labels for each region. Various graphical models have been employed for semantic segmentation \cite{lin2016scribblesup,zhang2013representative}. (c) Object localization methods, which are based on discriminative annotations, where seeds are produced by CNNs and Class Activation Mapping (CAMs) and then improved using localization enhancement methods \cite{gao2020robust,wang2018weakly}. (d) In self-supervised methods, image-level annotations are utilized to generate provisional segmentation masks, and pixel-level segmentations are learned from these masks. Some approaches iterate between fine predictions \cite{hinton2011transforming} or employ CAMs \cite{selvaraju2017grad} and saliency maps \cite{simonyan2013deep} as initial seeds.

\subsection*{Interpretation of Dynamic Routing} 
So far, various methods have been introduced in the literature to interpret CNNs, which can broadly be categorized into two main groups. The first category is architecture-agnostic, utilizing multiple providers simultaneously. Two useful methods in this category are Integrated Gradients \cite{sundararajan2017axiomatic} and SmoothGrad \cite{smilkov2017smoothgrad}. The second category is based on specific model layers, such as Guided Backpropagation \cite{cao2020dual}, and Grad-CAM \cite{chlebus2018automatic}. Grad-CAM is a technique that calculates feature maps by taking the derivatives of each class with respect to the feature maps of the last convolutional layer. Additionally, several methods have been developed based on Grad-CAM, including Grad-CAM++ which leverages second-order gradients \cite{chattopadhay2018grad}, and Score-CAM which utilizes scale activations to change the image \cite{wang2020score}. Other methods have been developed such as SS-CAM, a combination of SmoothGrad and Score-CAM \cite{wang2020ss}, and Full-GRAD, which is the summation of gradients from all layers \cite{srinivas2019full}. The majority of these interpretation methods cannot be easily applied to Capsule Networks due to their iterative routing mechanism. Furthermore, there is currently no interpretation method specifically designed for Capsules. However, there are a few architecture-agnostic methods, such as SmoothGrad \cite{smilkov2017smoothgrad}, that can be directly generalized to Capsule Networks. These methods only require the gradients of the output with respect to the input.

\subsection*{Histopathological Semantic Segmentation} This technique plays a crucial role in enabling medical professionals to accurately identify and diagnose diseased tissues. Various approaches can be employed for histopathological semantic segmentation. For instance, CNNs based on super-pixels have proven to be effective in segmenting the nucleus and tissues, as demonstrated in studies by \cite{turkki2016antibody,xu2016deep,chan2019histosegnet}. Sliding patch-based CNN methods, focusing on the segmentation of glands, cells, and mitosis, have been utilized in research by \cite{kainz2015semantic,shkolyar2015automatic,malon2013classification}. Additionally, other techniques leverage MIL based on weakly supervised learning, as explored by \cite{wang2018weakly,xu2014weakly}.  

\section*{Dataset}
The ADP datasets used in this study contain 100 glass slides selected from a pool of 500 slides. These slides were stained with Hematoxylin and Eosin (H\&E). The selection criteria for these slides were based on various features, including consistent tissue thickness, the absence of artifacts such as bubbles and tissue folding/crushing/cracks, the inclusion of diverse tissues from most organs in the body, and the presence of different diseases. Digitalization of the slides was carried out using a Huron TissueScope LE1.2 WSI scanner, resulting in images with dimensions of $272\times272$ pixels and a resolution of 0.0625 $\mu$m \cite{hosseini2019atlas}. 

\begin{figure}[t!]
\centering
\includegraphics[width=0.75\linewidth]{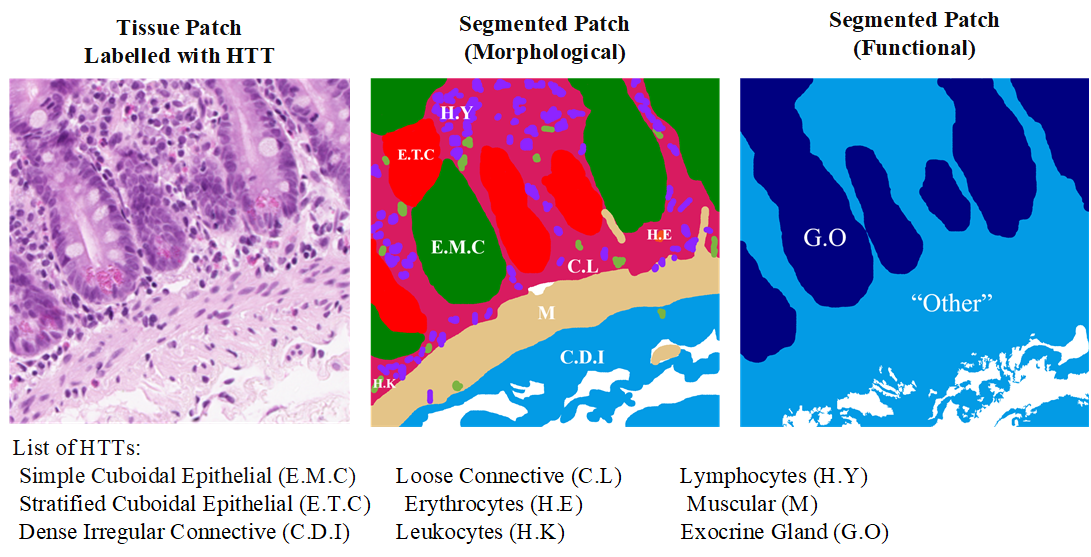}
\caption{\footnotesize The proposed methodology utilizes training based on annotations of histological tissue segments and forecasts both the morphological and functional types of tissue at the granular level of individual pixels.}
\label{figure:fig1}
\end{figure}

\begin{figure}[t!]
\centering
  \includegraphics[scale=0.15]{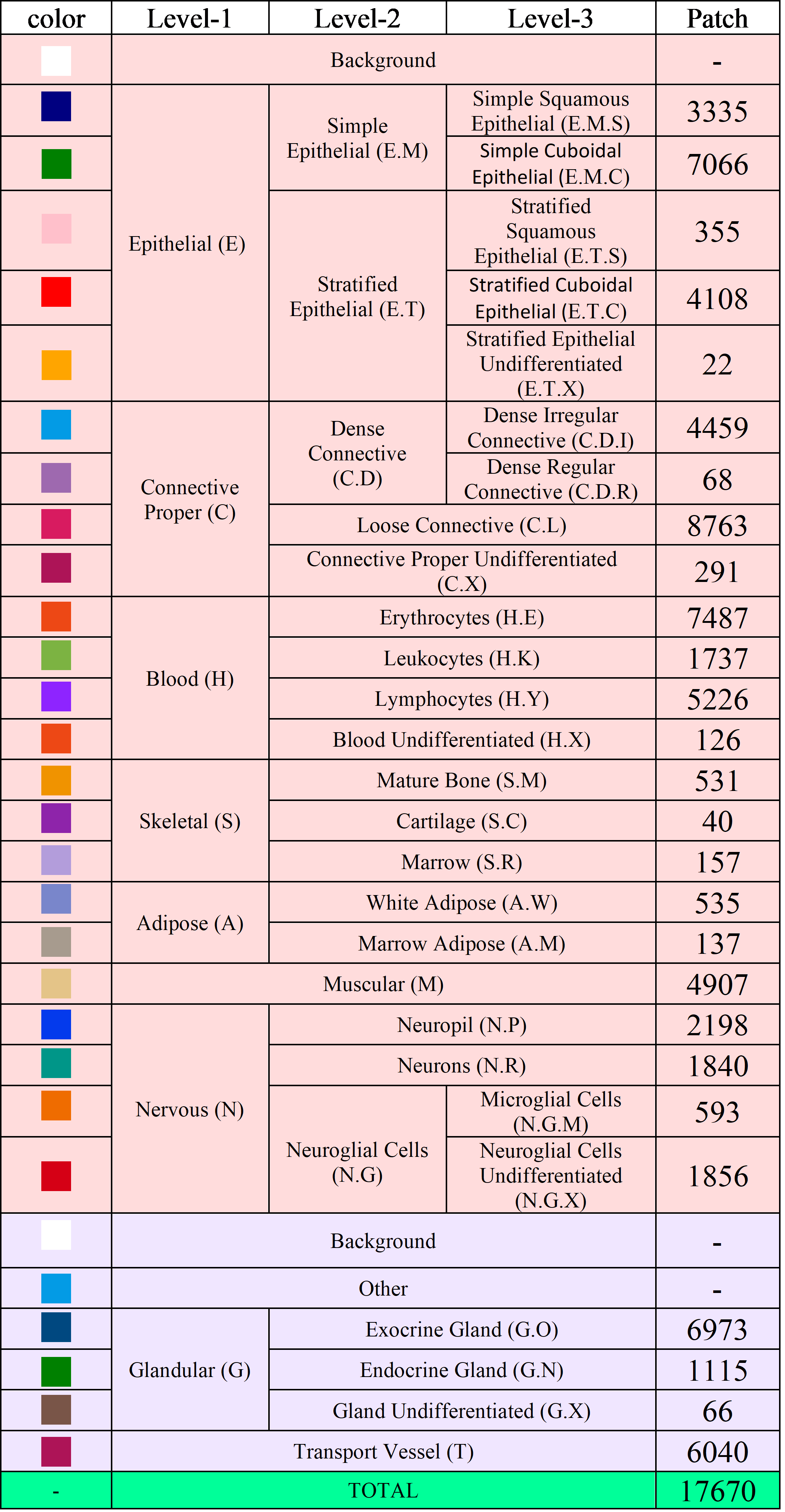}
  \caption{\footnotesize The proposed Atlas database employs a structured classification of histological tissue types for guided annotation. This tissue classification system is organized into three tiers, progressing from the broadest category at the top to the most detailed at the bottom.\label{figure:fig2}}
\end{figure}

\indent    To label the patches within these datasets, a Hierarchical Tissue Taxonomy was employed. In histology, two practical approaches are commonly used for tissue classification: morphological assortment, which focuses on studying tissue structure, and functional assortment, which investigates the function of organs such as glandular or vascular structures. The assortment categories have been color-coded, as illustrated in Fig. \ref{figure:fig1}. To separate tissues based on taxonomy, three patch levels were used. The first patch level consists of nine tissues, two of which are functional, while the remaining are morphological categories, providing more specific sub-types. These tissues are further divided into the second and third patch levels. The third one contains more detailed tissue distinctions, encompassing 27 types, including 23 morphological and 4 functional categories. Fig. \ref{figure:fig2} provides illustrations of HTTs at the patch and pixel levels, allowing visualization of the morphological and functional tissue concepts. In this work, the functional category also includes a “Background” designation for non-tissue and “Other”  for non-functional tissue regions. The “Background” part is also included in the morphological assortment. Lastly, 43 manually segmented patches are used to quantitatively fine-tune our method.

\section*{Methodology}
\begin{figure*}[t!]
\centering
  \includegraphics[width=\linewidth]{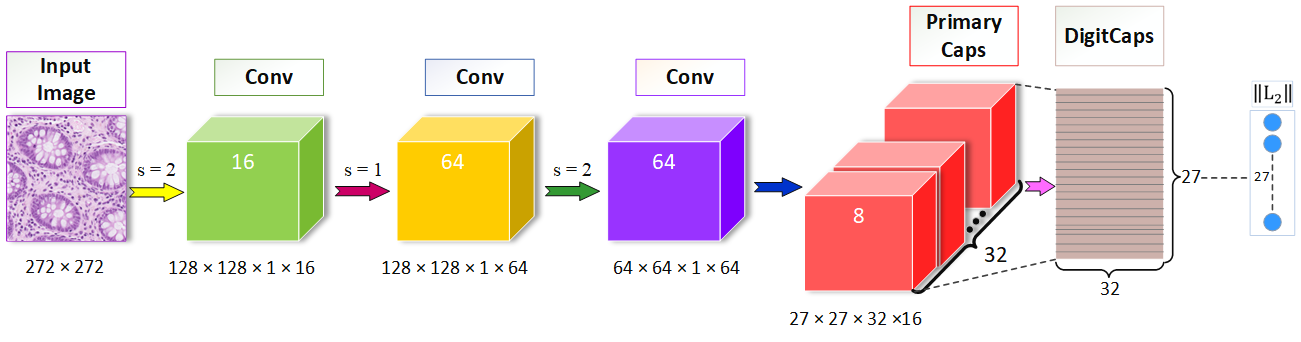}
  \caption{\footnotesize Proposed HistoSegCap architecture.}
  \label{figure:fig3}
\end{figure*}

In this section, we present the proposed HistoSegCap methodology. To create semantically segmented results, each patch is progressed through the following stages: (1) patch-level classification to predict potential tissue classifications for the input patch, (2) pixel-level reconstruction stage to predict blurred spatial locations for each detected class at the pixel level, (3) pixel-level segmentation for the creation of pixel-level activation maps, (4) non-relevant parts elimination stage to detect non-functional and non-tissue parts, and finally (5) fusion interpretation stage to combine the information of the reconstruction and activation maps . A visual representation of the HistoSegCap architecture is presented in Fig. \ref{figure:fig3}, providing a comprehensive overview of the process. 

\subsection*{Patch-Level Classification}

As previously mentioned, the primary objective of this paper is to develop a cutting-edge Capsule Networks architecture for histopathology semantic segmentation. To achieve this goal, we begin by outlining the fundamental properties of Capsule Networks. Subsequently, we discuss the proposed HistoSegCap architecture. In Capsule Networks, numerous neurons are present, and their activity vectors play a crucial role in determining the position and orientation of the Capsule. The magnitude of this vector serves as an indicator of the probability that a specific object will be represented by it. To illustrate the concept of probability, it is essential to note that the length of this network is restricted by the squashing function, which confines it within the range of $[0,1]$ \cite{sabour2017dynamic}. The mathematical expression for this function is given by
\begin{equation}
\textbf{v}_j=\frac{{\left.\|\textbf{s}_j\|\right.}^2}{1+{\left.\|\textbf{s}_j\|\right.}^2}\frac{\textbf{s}_j}{\left.\|\textbf{s}_j\|\right.},
\end{equation}
where for the Capsule $j$, the output vector is ${\textbf{v}}_j$, and the total input is ${\textbf{s}}_j$. All Capsules, with the exception of the first one, obtain their ${\textbf{s}}_j\ $from ${\hat{\textbf{u}}}_{j|i}$ which is the weighted sum of the prediction vector of the lower Capsules, i.e.,
\begin{equation}
{\boldsymbol{\mathrm{\textbf{s}}}}_j=\sum_i{}c_{ij}{\hat{\boldsymbol{\mathrm{\textbf{u}}}}}_{j\mid i},
\end{equation}
where ${\hat{\textbf{u}}}_{j|i}$ is determined as the multiplication of the lower Capsule output ${\boldsymbol{\mathrm{\textbf{u}}}}_i$, and the weight matrix ${\textbf{W}_{ij}}$,
\begin{equation}
{\hat{\boldsymbol{\mathrm{\textbf{u}}}}}_{j\mid i}={\boldsymbol{\mathrm{W}}}_{ij}{\boldsymbol{\mathrm{\textbf{u}}}}_i.
\end{equation}
The iterative dynamic routing process plays a crucial role in determining the coupling coefficients, denoted by $c_{ij}$, in Capsule Networks. To ensure that the coupling probabilities between Capsule $i$ and the higher layers sum up to $1$, the SoftMax function is employed. This constraint is achieved by utilizing $b_{ij}$, which represents the previous probability of coupling between the primary Capsule $i$ and the subsequent Capsule $j$. More precisely,
\begin{equation}
c_{ij}=\frac{{\mathrm{exp} \left(b_{ij}\right)\ }}{\sum_k{\ }\mathrm{exp}\left(b_{ik}\right)}.
\end{equation}
During the learning process, the previous probability is continuously updated with the weights, and its location is determined by the types and locations of the two Capsules involved, rather than being dependent on the current input image. The primary coupling coefficient, $c_{ij},$ is then updated based on the consistency between the output of the previous high-level capsule, ${\textbf{v}}_j$, and the prediction  ${\hat{\textbf{u}}}_{j|i}$ to be made by Capsule $i$. This consistency is determined by a scalar value $a_{ij}={\textbf{v}}_j.{\hat{\textbf{u}}}_{j|i}$, which is added to ${b}_{ij}$ before performing a new calculation for $c_{ij}$.

Further, we provide a detailed description of our proposed architecture. The HistoSegCap architecture, as depicted in Fig. \ref{figure:fig3}, consists of various components, each of which will be comprehensively explained in the following sections.

\subsubsection*{Feature Maps} The HistoSegCap model requires a sizeable input image of $272\times272$ pixels. This image is then passed through three convolutional layers to reduce input image dimensionality. This downsampling helps in minimizing the computational complexity of subsequent layers and extracting higher-level features. By using convolutional layers in Capsule Networks, we can effectively reduce spatial dimensions while preserving important features. A $3\times3$ convolution kernel is considered for all layers of convolution.

\subsubsection*{Primary Capsules} In the primary Capsule stage, the model generates new Capsules using a $2\times2$ stride and $9\times9$ convolution kernels. These layers are composed of 32 distinct capsules, each containing 8 dimensions and feature maps measuring $27\times27$. Furthermore, the model incorporates dynamic routing between the primary Capsules and the subsequent layers, referred to as DigitCaps. This process facilitates the efficient flow of information between Capsules, ultimately enabling the model to accurately classify and segment images. 

\subsubsection*{Digit Capsules} In these layers of the HistoSegCap model, the dynamic routing theory is utilized to generate 27 labels. This process involves using various weights to generate labels from the 32-dimensional Capsules for each class of digits. The dynamic routing technique utilized in this stage is a crucial element in the model's ability to accurately classify and segment images. By dynamically adjusting the weights of the connections between Capsules, the model can effectively capture the complex relationships between different features in the image. This process results in enhanced accuracy and robustness, allowing the model to effectively handle complex image characteristics and achieve superior performance.

\subsubsection*{Reconstruction} 
\begin{figure*}[t!]
\centering
  \includegraphics[width=\linewidth]{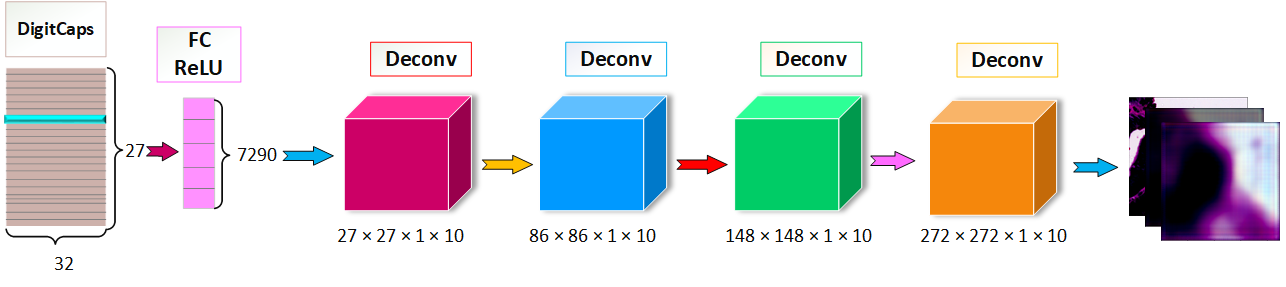}
  \caption{\footnotesize Proposed reconstruction architecture.}
  \label{figure:fig4}
\end{figure*}

The reconstruction layer operates by taking the outputs from the DigitCaps layer and endeavors to reconstruct the original input image. The process involves solely those Capsules estimated by the network, whose labels are present in the input image. To recognize labels in an image, the network compares the length of each label's Capsule against the predefined threshold levels. Once the labels for each image are determined, the values of each label in the image are extracted separately from the DigitCaps layer, and the outputs are meticulously directed to a decoder responsible for reconstructing the input image. As illustrated in Fig. \ref{figure:fig4}, the decoder is composed of four deconvolutional layers to create the reconstruction maps, i.e., ${\textbf{M}}_j^{rec}$, for each label in the input image, represented by $j\in J$. Note that ${\textbf{M}}_j^{rec}$ plays a crucial role in reconstructing images associated with each label, providing a visual representation that illustrates the position and range specific to that label. This strategic approach ensures higher accuracy and efficiency in classification and interpretation processes.

\subsubsection*{Loss Function} To attain effective model training, one must give due consideration to the cumulative influence of both reconstruction loss and margin loss. These two components play a crucial role in shaping the process and ensuring the model’s overall performance. Through addition of the reconstruction loss and the margin loss, as outlined below, a comprehensive and well-balanced training approach can be achieved, maximizing the model’s learning capabilities and enhancing its overall accuracy and effectiveness:
\begin{itemize}
\item \textbf{Reconstruction Loss:} As previously mentioned in the reconstruction section, the process of identifying labels within the image is accomplished by applying the normalized digit Capsule data to a predetermined threshold. Following this, the output of the identified labels is passed through the decoder to reconstruct the image's label. This process results in a set of ${\textbf{M}}_j^{rec}$ identified representations, whose combination should reconstruct the input image,
\begin{equation}
\hat{\textbf{{I}}}= \sum_{j\in J} \textbf{M}_j^{rec},
\end{equation}
by following these steps, we can achieve an output that successfully reconstructs all the labels within the image. Thus, the reconstruction loss can be computed as follows
\begin{equation}
L^{rec}= {\|{\textbf{I}}-{\hat{\textbf{{I}}}}\|}^2.
\end{equation}
It is essential to emphasize that the reconstruction loss typically exhibits a substantial value. Therefore, to combine it with the margin loss, the reconstruction loss is multiplied by a small factor, denoted by $\alpha$. To attain superior outcomes, the $\alpha$ value decreases exponentially after each cycle.
\item \textbf{Margin Loss:} Our objective is for the Capsule associated with class $j$ to possess a lengthy instantiation vector only when that particular label is present within the image. To accommodate scenarios where multiple labels are present, we implement a distinct margin loss, denoted as ${L_j^{mrg}}$, for each label Capsule. This guarantees that each label is adequately taken into account and contributes to the overall loss calculation,
\begin{equation}
{\small{L_j^{mrg}}  = T_j\max{\left(0,m^+-\left.\|{\boldsymbol{\mathrm{v}}}_j\|\right.\right)}^2 +\lambda \left(1-T_j\right)\max{\left(0,\left.||{\boldsymbol{\mathrm{v}}}_j\|\right.-m^-\right)}^2.}
\end{equation}
During the initial learning process, the loss for each missing class is down-weighted by $\lambda$ to prevent a reduction in the activity vectors of the label Capsules. The values of $m^+$ and $m^-$ are set to 0.9 and 0.1, respectively. Additionally, when a class of $j$ is present, the value of $T_j$ is set to 1, otherwise it is 0.
\end{itemize}

\subsection*{Pixel-Level Reconstruction}
When utilizing reconstruction as a regularization technique, it proves specific for certain datasets like MNIST. Nonetheless, it may encounter challenges when applied to colorful datasets such as CIFAR-10 and ADP. These challenges include computational complexity, loss of fine-grained details, and limitations in capturing dataset diversity. As a result, the output of the reconstruction method for these datasets may exhibit some blurriness and a lack of perfect clarity. Despite these limitations, it still yields several advantages, such as improved target localization, reduced noise, enhanced generalization, improved interpretability, and robustness against dataset diversity. It is crucial to emphasize that while the method may not yield entirely evident outcomes, it remains valuable in achieving more accurate target localizations. Through the preservation of significant details and noise reduction, it contributes to enhancing the overall output quality. Furthermore, the interpretability of the model is enhanced through the reconstruction method. By analyzing the reconstructed images, it becomes easier to grasp the features and patterns that the model considers necessary for target localization. This insight be valuable in gaining a deeper comprehension of the model's decision-making process. 

\subsection*{Pixel-Level Segmentation}
CNNs provide a variety of efficient WSSS methods at the patch level. Nevertheless, no such techniques have been specifically documented for Capsule Networks. This paper proposes a novel approach by incorporating gradient-based methods within a Capsule Network. These methods involve the analysis of neural network gradients with respect to their inputs. The gradient-based methods aim to comprehend the significance or relevance of different input features or regions on the network's output. In particular, these interpretation methods produce a saliency map for each class $j$, denoted by ${\textbf{M}}_j^{sal}\mathrm{(}{\textbf{I}}\mathrm{)}$, by computing the derivative of the target class score ${{S_j}}\mathrm{(}{\textbf{I}}\mathrm{)}$ to the input image ${\textbf{I}}$, for each pixel $(x,y)$, i.e., 
\begin{equation}
{\textbf{M}}_j{_{(x,y)}}^{sal}\mathrm{(}{\textbf{I}}\mathrm{)}=\mathrm{\partial }S_j\mathrm{(}{\textbf{I}}\mathrm{)/}\mathrm{\partial }{\textbf{I}_{(x,y)}}.
\end{equation}
In the HistoSegCap model, the class scores are actually the norm layers' outputs. In situations where the derivative of ${{S_j}}\mathrm{(}{\textbf{I}}\mathrm{)}$ is evaluated at smaller scales, it may display sharp oscillations, leading to a sensitivity map that appears noisy and contains sampling variations in partial derivatives that lack meaningful interpretations. SmoothGrad can be used in such conditions to mitigate these issues \cite{smilkov2017smoothgrad}. By generating multiple perturbed versions of the input image and averaging the gradients over these perturbations, SmoothGrad smooths out sharp oscillations. This helps to reduce noise in the sensitivity maps. The SmoothGrad technique can be mathematically expressed in the following fashion
\begin{equation}
{\textbf{M}}_j^{smooth}=\frac{1}{N}\sum^N_{n=1}{\ }{{\textbf{M}}}_j^{sal}\left({\textbf{I}}+\textbf{Z}_n)\right),
\end{equation}
where $N$ represents the number of samples and $\textbf{Z}_n \sim \mathcal{N}(0,{\sigma}^2)$ denotes an additive Gaussian noise with zero mean and variance of ${\sigma }^2$. As seen, the SmoothGrad technique is controlled by two hyperparameters $\sigma$ and $N$. Adjusting these hyperparameter values can have a notable impact on the final output. By employing fine-tuned hand-segmented patches, the values of the optimal hyperparameters were determined to be $N=40$ and $\sigma = 0.15$.

\subsection*{Non-Relevant Parts Elimination}
Creating artificial “Background” labels for both morphological and functional modes, as well as “Other” activations for the functional mode, is imperative in the ADP database due to the lack of labels for non-functional and non-tissue components. These activation maps play a crucial role in preventing predictions when valid pixel classes are unavailable in the ADP database
\begin{itemize}
\item \textbf{``Background'' Elimination:} In the WSIs, the region with high white illumination values encompasses both background and transparently stained tissues, such as white/brown adipose, glandular, and transport vessels. Therefore, it is crucial to have an activation map specifically for the background to distinguish these elements. The process of background activation involves applying a scaled-and-shifted Sigmoid to the mean of the RGB image, denoted as $\overline{\textbf{I}}$, resulting in the generation of white illumination images. Following this, the activations of the transparent staining class need to be subtracted. Finally, a 2D Gaussian blur is applied to the resulting image to reduce the resolution of the prediction. As a result, the background activation mask can be computed as follows
\begin{eqnarray}
{\textbf{M}}^B_{(x,y)} &\leftarrow& \frac{0.75}{1+\mathrm{exp}[-4({\overline{\textbf{I}}}_{(x,y)}-240)]}\\
{\textbf{M}}^{B_{morph}} &\leftarrow& \left({\textbf{M}}^B-\max\left({\textbf{M}}^{F}_{\mathrm{A.W}}\right)\right)*H_{0,2}\\
{\textbf{M}}^{B_{func}} &\leftarrow& \left({\textbf{M}}^B-\max\left(\textbf{M}^{F}_{\mathrm{G.O}},{\textbf{M}}^{F}_{\mathrm{G.N}},{\textbf{M}}^{F}_{\mathrm{T}}\right)\right)*H_{0,2}. 
\end{eqnarray}
\item \textbf{``Other'' Activation:} Pixels associated with non-functional tissues, distinct from the background, should exhibit low activations for both background and all other functional tissues in the functional mode. To achieve this, the 2D maximum must be computed for other functional types, the background, and the adipose activations based on morphological data. Finally, the probability map must be scaled by 0.05 after subtracting one from it, i.e., 
\begin{equation}
 \centering {\textbf{M}}^{\mathrm{O\ }}\leftarrow 0.05\left[1-\max\left({\left\{{\textbf{M}_j}^{F}\right\}}_{j\in J_{func}},{\textbf{M}}^{B_{func}},{\textbf{M}}^{F}_A\right)\right].  
\end{equation}
\end{itemize}

\begin{figure}[t!]
\centering
  \includegraphics[width=1\linewidth]{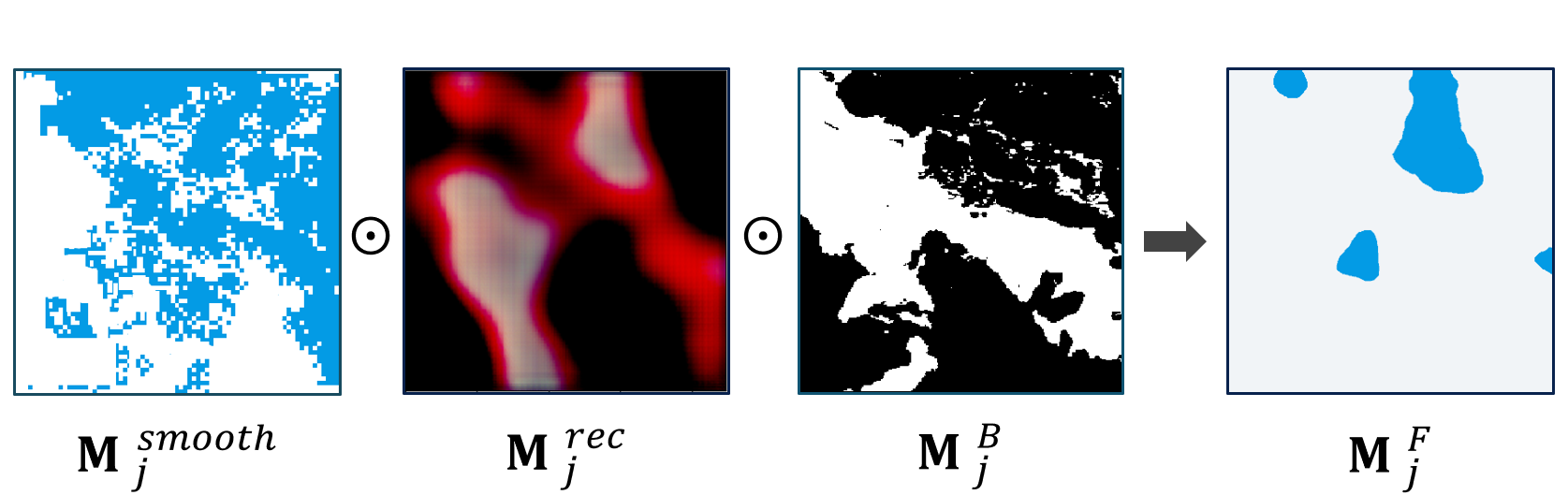}
  \caption{\footnotesize This figure demonstrates the application of image reconstruction technique in generating output results for a specific tissue type. The process enhances the clarity and detail of the tissue image, thereby improving the accuracy of the results.}
  \label{figure:fign}
\end{figure}

\subsection*{Fusion Interpretation Approach}
Now we are ready to obtain the model output by synthesizing the results of two interpretation techniques and removing the background parts. For each label in the image, both the SmoothGrad and the Reconstruction methods assign a value ranging from 0 to 1 to each pixel. Achieving superior results can be accomplished by integrating information from ${\textbf{M}}_j^{rec}$ and ${\textbf{M}}_j^{smooth}$. This process reinforces the common areas between the reconstruction and the SmoothGrad results to improve diagnostic quality, Fig. \ref{figure:fign}. Finally, by removing the background from the result, the morphological output is,
\begin{equation}
\textbf{M}_j^F={\textbf{M}_j^{rec}}\odot {\textbf{M}_j^{smooth}}\odot {\textbf{M}^{B_{morph}}},             \hspace{0.5cm}\forall j\in J,
\end{equation}
where $\odot$ is Hadamard product. In functional mode, it is also necessary to remove the ``Other'' part from the result. To generate the final output, encompassing each pixel $(x,y)$, we calculate
\begin{equation}
\textbf{Q}_{(x,y)}=\operatorname*{arg\,max}_{j\in{J}} \ \ \  (\textbf{M}_{j_{(x,y)}}^{F}),
\end{equation}
where $\textbf{Q}_{(x,y)}$ represents the predicted segmentation mask for all semantic classes within an image. This process enables us to identify the values corresponding to the existing labels and generate a comprehensive image that incorporates all the desired labels. By employing this meticulous approach, we ensure that the final output accurately represents the intended labels and meets the highest standards of precision and quality.

\section*{Results}
The subsequent section presents a comprehensive examination of the performance of the HistoSegCap network, with a specific emphasis on quantitative evaluation. This evaluation is initially carried out on hand-segmented images from the ADP dataset, providing a measure of the network's effectiveness using the mean Intersection-over-Union (mIoU) criterion as will be described shortly. Subsequently, a comparative analysis is conducted, contrasting the HistoSegCap network with existing WSSS methodologies \cite{kolesnikov2016seed,huang2018weakly,chan2019histosegnet}. The objective of this comparison is to provide a comprehensive perspective on the performance and precision of HistoSegCap. To conduct these experiments, the PyTorch framework was employed for both training and testing phases with an NVIDIA RTX 3090 GPU.

\subsection*{Performance Assessment}
\begin{figure}[t!]
\centering
  \includegraphics[width=1\linewidth]{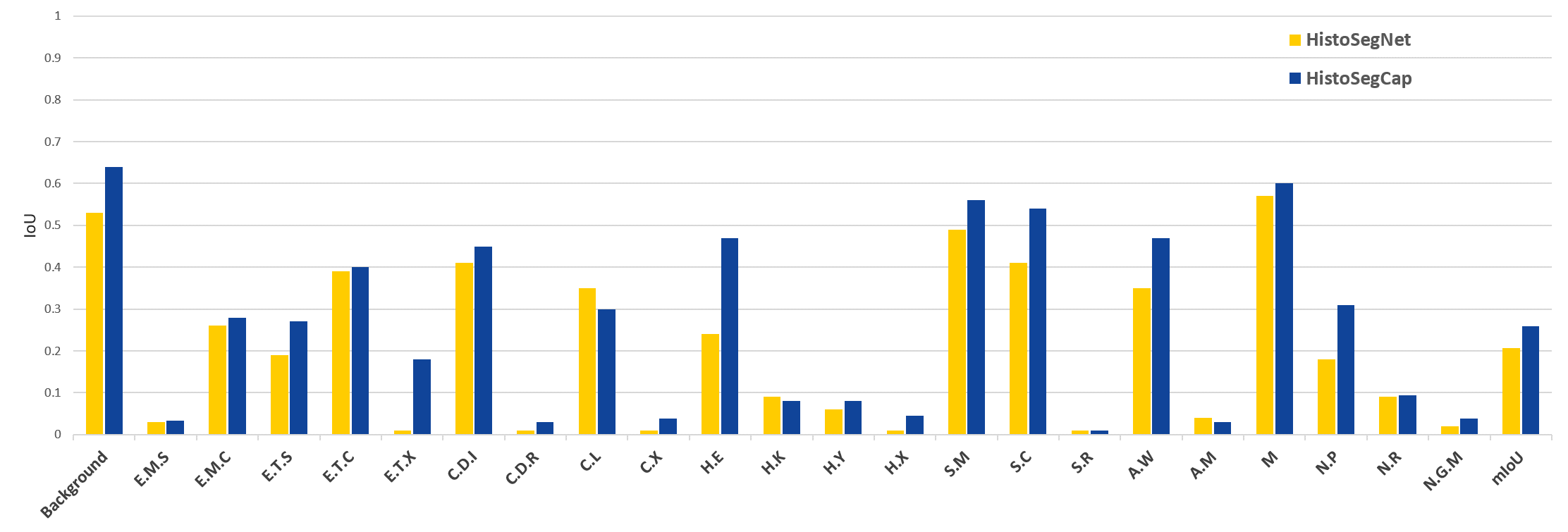}
  \caption{\footnotesize Comparison of Intersection over Union (IoU) for predicted and ground truth segmentations in the tuning set for different morphological types using both the proposed HistoSegCap and HistoSegNet\cite{chan2019histosegnet} models.}
  \label{figure:fig8}
\end{figure}

\begin{figure}[t!]
\centering
  \includegraphics[width=0.7\linewidth]{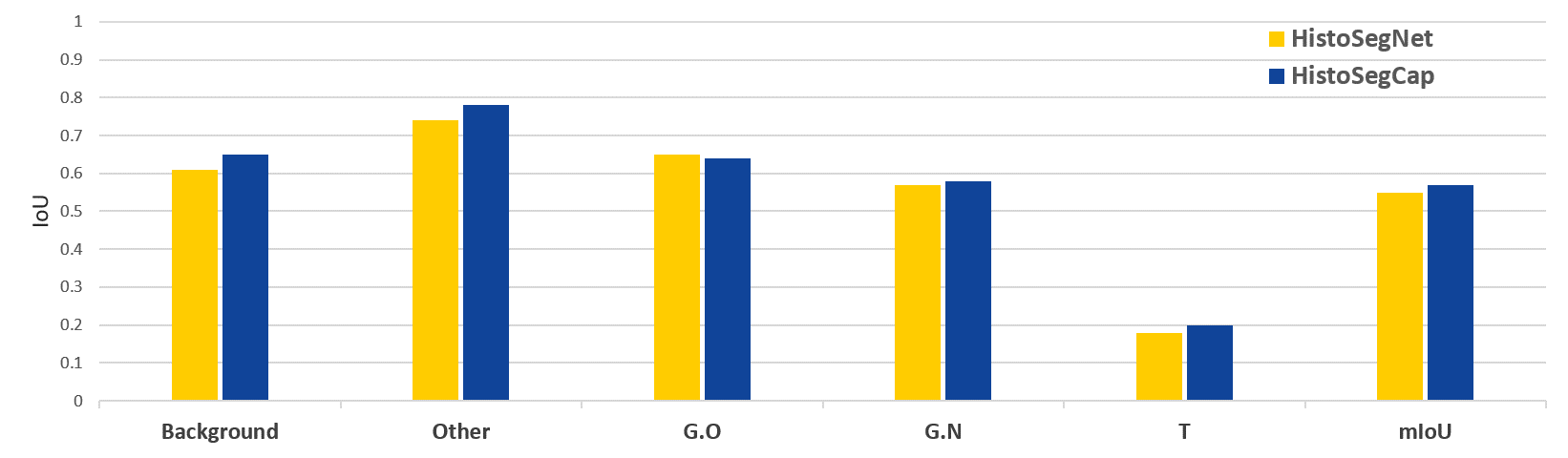}
  \caption{\footnotesize Comparison of IoU for predicted and ground truth segmentations in the tuning set for different functional types using both the proposed HistoSegCap and HistoSegNet\cite{chan2019histosegnet} models.}
  \label{figure:fig9}
\end{figure}

A set of 43 meticulously hand-segmented images from the ADP database has been selected. Each image was segmented at the pixel level by a skilled pathologist. The performance evaluation of the HistoSegCap architecture and the comparative analysis with other methods were carried out using the mIoU metric. This criterion enables the evaluation of effectiveness in the segmentation result.  The mIoU assesses the accuracy of segmentation algorithms by quantifying the overlap between the ground truth $(\textbf{G})$ and predicted segmentations $(\textbf{P})$ at the pixel level as follows
\begin{equation}
\mathrm{mIoU}=\frac{1}{|J|} \sum_{j \in J} \frac{\left|\mathbf{P}_j \cap \mathbf{G}_j\right|}{\left|\mathbf{P}_j \cup \mathbf{G}_j\right|} .
\end{equation}
To be more precise, the mIoU metric is computed by dividing the intersection of the predicted and ground truth regions by their union and then averaging this value across all $J$ classes. The mIoU criterion was computed for both morphological and functional tissues using Eq. (16). The analysis of Fig.~\ref{figure:fig8} and Fig. \ref{figure:fig9}, which represent morphological and functional types respectively, reveals that HistoSegCap demonstrates superior performance on the tuning set for functional types with a mIoU of 0.5675, compared to its performance on morphological types where the mIoU is 0.2587. In the morphological mode, the Histological Tissue Types (HTTs) that show the best performance are Mature Bone (S.M) and Skeletal Muscular (M), while the HTTs with the least performance are those with fewer ground-truth examples, such as C.D.R. On the other hand, the performance is more stable in the functional mode, with the lowest performance observed for Transport Vessel (T).

\begin{figure}[t!]
\centering
  \includegraphics[width=0.45\linewidth]{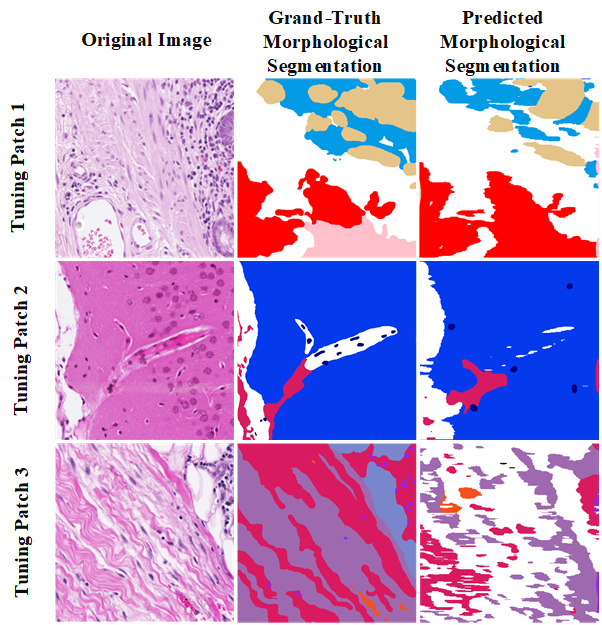}
    \caption{\footnotesize A visual analysis was performed on selected image regions from the dataset utilized for model optimization. These segmented areas were juxtaposed with the ground-truth segmentations of morphological tissues. For reference on the color-coding system used in the segmentations, please refer to Fig. \ref{figure:fig2}.}
  \label{figure:fig5}
\end{figure}

\begin{figure}[t!]
\centering
  \includegraphics[width=0.45\linewidth]{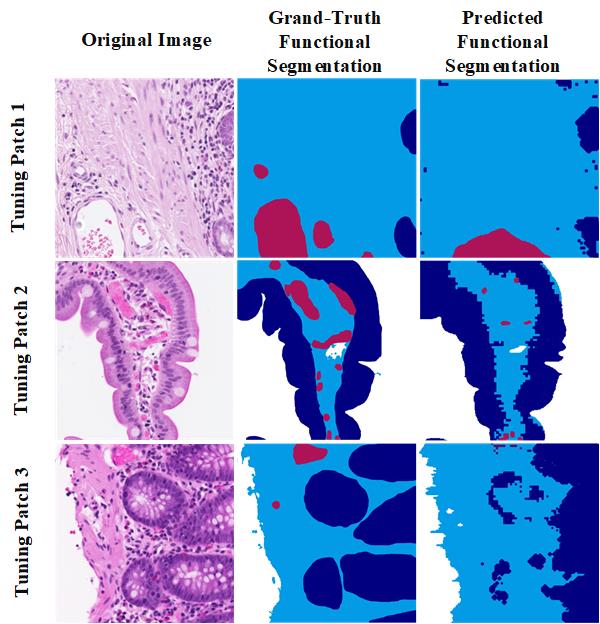}
  \caption{\footnotesize The segmented regions obtained through the proposed network were meticulously compared with the well-established ground-truth segmentations for various functional tissues.}
  \label{figure:fig6}
\end{figure}

Fig.~\ref{figure:fig5} and Fig.~\ref{figure:fig6} provide visual confirmation of the network's proficiency in semantic segmentation for morphological and functional tissues, respectively. The segmented images exhibit strong concordance with the ground truth images, capturing even small textural details delineated properly. Overall, these quantitative and qualitative results validate the ability of the HistoSegCAP network to perform excellent tissue segmentation.


\subsection*{Comparison with State-of-the-Art WSSS}
\begin{figure}[t!]
\centering
\includegraphics[width=0.9\linewidth]{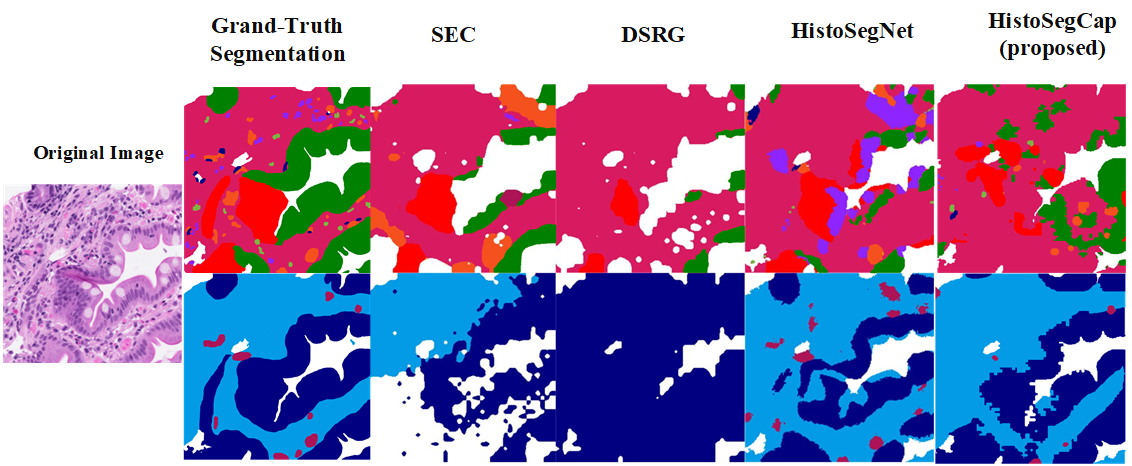} 
\caption{\footnotesize A patch segmentation comparison between the HistoSegCap model and various WSSS methods (SEC, DSRG, and HistoSegNet)}
\label{figure:fig7}
\end{figure}

To assess the effectiveness of the proposed HistoSegCap model, a comparative analysis is performed against several other state-of-the-art WSSS networks, including SEC \cite{kolesnikov2016seed}, DSRG \cite{huang2018weakly}, HistoSegNet \cite{chan2019histosegnet}, SEAM \cite{wang2020self}, MPS-PDA \cite{han2022multi}, and Histo-Puzzle \cite{ma2023histo}. The training of the HistoSegNet framework involves utilizing CNNs over a span of 80 cycles, employing a cyclical learning rate and a batch size of 16. Furthermore, enhancements are made to the Fully Convolutional Network (FCN) components of both SEC and DSRG through refinements using the ADP database. SEC and DSRG employ a gradual reduction strategy for their learning rates, with a decay rate of 0.5 every four cycles, commencing from $10^{-4}$, spanning over 16 cycles. A comprehensive comparison of these state-of-the-art architectures with our proposed method, based on the mIoU criterion, can be found in Table \ref{table:tab1}, providing in-depth insights. Moreover, Fig. \ref{figure:fig7} presents a comparative depiction of the segmented results for the same image using different methods.
\begin{table}[t!]
\centering
\begin{tabular}{|l|l|l|}
\hline
\textbf{Methods} & \textbf{morphological} & \textbf{functional}\\
\hline
SEC \cite{kolesnikov2016seed}&  0.1628&0.3225  \\ 
\hline
DSRG \cite{huang2018weakly}& 0.1375 & 0.4732  \\
\hline
HistoSegNet \cite{chan2019histosegnet}& 0.2206 &0.5505  \\
\hline
SEAM \cite{wang2020self}& 0.2539 &0.5051  \\
\hline
MPS-PDA \cite{han2022multi}& 0.0939 &0.3058  \\
\hline
Histo-Puzzle \cite{ma2023histo}& 0.2223 &0.5624  \\
\hline
HistoSegCap (proposed)& \textbf{0.2587} & \textbf{0.5674}  \\
\hline
\end{tabular}
\caption{\footnotesize Quantitative comparison of WSSS methods using the mIoU metric. \label{table:tab1}}
\end{table}

Further investigation and a general assessment of the leading CNN and Capsule models are also conducted. The accuracy of the model can be calculated by adding up the count of the true positives and true negatives, and then dividing the result by the total number of outcomes, i.e.,
\begin{equation}
\text {\text{Accuracy}}=\frac{(T P+T N)}{(T P+T N+F P+F N)}\times100 \%,
\end{equation}
where True Positive (TP) refers to the count of accurately predicted positive outcomes, indicating the instances where the model correctly identifies positive results. Conversely, True Negative (TN) represents the number of accurately predicted negative outcomes, demonstrating its precise identification of negative results. The accuracy calculation results for the two networks are presented in Table \ref{table:tab2}, clearly demonstrating the superiority of the HistoSegCap model.
\begin{table}[t!]
\centering
\begin{tabular}{|l|l|}
\hline
\textbf{Methods} & \textbf{Accuracy}\\
\hline
HistoSegNet \cite{chan2019histosegnet}& 94.3\%  \\ 
\hline
HistoSegCap (proposed)&95.2\%    \\
\hline
\end{tabular}
\caption{\footnotesize An accuracy comparison between HistoSegNet \cite{chan2019histosegnet} and the proposed HistoSegCap network. \label{table:tab2}}
\end{table}

In summary, the findings suggest that the HistoSegCap model demonstrates effectiveness as a CAD tool for classification and semantic segmentation. Moreover, the model's visual assistance capabilities enhance pathologists' ability to thoroughly analyze whole slide images, leading to a more effective examination.
\subsection*{Detecting Diseased Tissues}
\begin{figure}[t!]
\centering
\includegraphics[width=1\linewidth]{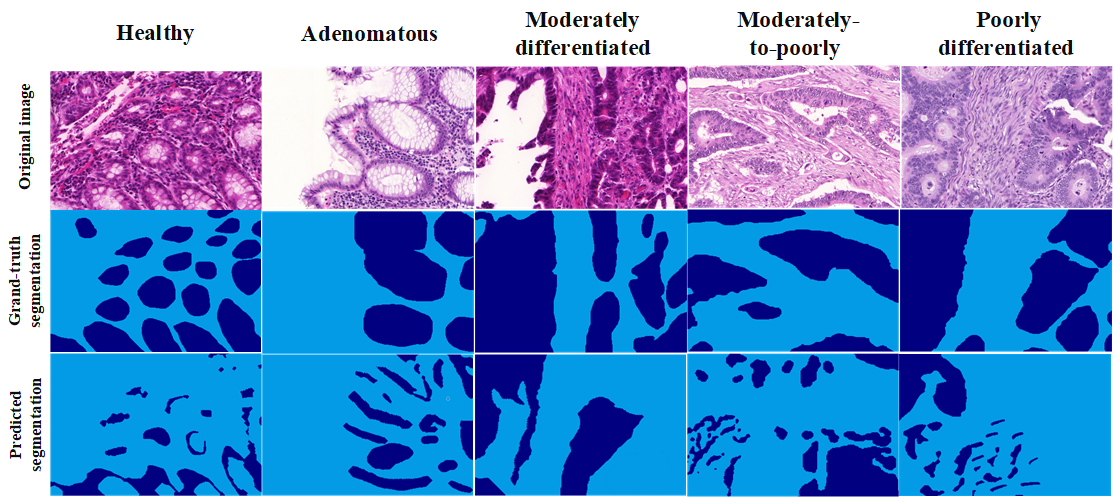} 
\caption{\footnotesize A patch segmentation performance of HistoSegCap on Warwick-QU dataset based on different tumor grades.}
\label{figure:fig7n}
\end{figure}
In this section, we evaluate the proposed HistoSegCAP model using the Warwick-QU dataset, which consists of both healthy and diseased tissues \cite{sirinukunwattana2017gland}. Although the model is initially trained on the ADP dataset, which primarily contains healthy tissues, it can still be utilized to detect diseased tissues as well. The Warwick-QU dataset comprises 165 H\&E-stained histology images of colon glands with varying cancer grades. These images are manually annotated for gland segmentation and classification. The results of our approach demonstrate that by training the model on healthy tissues and learning their segmentation, it can effectively detect diseased tissues as well.

Since the Warwick-QU dataset only includes two classes (glandular or non-glandular), HistoSegCAP is applied in functional mode to predict only  “G.O”and “Other” labels. Additionally, to align the input images with the model, the images are down-sampled to $272\times272$-pixel crops. Fig. \ref{figure:fig7n} showcases the qualitative performance of HistoSegCAP on selected images from the Warwick-QU dataset. Furthermore, Table \ref{table:tab3} provides a comprehensive quantitative evaluation of HistoSegCAP's performance in segmenting the Warwick-QU images at each tumor grade. The results indicate that HistoSegCAP's pixel-level predictions become progressively less confident and accurate as the tumor grade worsens.

\begin{table}[t!]
\centering
\begin{tabular}{|l|l|}
\hline
\textbf{Grade} & \textbf{“G.O”  IoU}\\
\hline
healthy& 0.5347  \\ 
\hline
 adenomatous & 0.3832    \\
 \hline
 moderately differentiated & 0.3514  \\
\hline
 moderately-to-poorly & 0.3302  \\
 \hline
 poorly differentiated & 0.3069  \\
  \hline
\end{tabular}
\caption{\footnotesize Segmentation performance on the Warwick-QU dataset: IoU for “G.O” across different tumor grades. \label{table:tab3}}
\end{table}

Overall, HistoSegCAP is shown to be capable of segmenting relevant tissues from slides scanned by different setups. Moreover, as the tumor grade deteriorates, the model's segmentations become less confident, exhibit less overlap, and appear more misshapen. This suggests that these segmentations can serve as predictive indicators of the level of disease in the tissue.

\section*{Conclusion}\label{sec:conclu}
In this paper, we presented a  pioneering approach to histopathological image analysis, leveraging WSSS in conjunction with Capsule Networks. The proposed novel model, named HistoSegCap, addressed the limitations inherent in existing CNN architectures, resulting in a significant performance enhancement in the semantic segmentation of WSIs. An integral aspect of our investigation involved the study of the impact of reconstruction layers on semantic segmentation. Our findings illuminated the potential of combining smoothGrad with reconstructed images, showcasing an augmented accuracy of label spatial detection. The HistoSegCap model was trained using the ADP dataset, which contains both morphological and functional HTTs. Moreover, the model was fine-tuned and evaluated on a hand-segmented subset of 43 images from ADP. Experimental results validated the superiority of the proposed HistoSegCap model when compared to existing semantic segmentation methods. In essence, the findings underscored the transformative capabilities of merging WSSS and Capsule Networks, offering a paradigm shift in histopathological image analysis. By facilitating patch-level representation of HTTs, the proposed model significantly enhanced the capability of CAD systems, aiding pathologists in the identification of diagnostically relevant regions. 

\bibliography{References}

\section*{Acknowledgments}
This work was partially supported by the Natural Sciences and Engineering Research Council (NARC) of Canada through the NARC Discovery Grant REPIN-2023-05654.

\section*{Data Availability}
The utilized dataset is publicly available through the link:\\
 {\url{https://www.dsp.utoronto.ca/projects/ADP/}}

\section*{Author Contributions Statement}
M.M. and S.SH. implemented the deep learning models, and performed the evaluations; M.M. and S.SH. drafted the manuscript jointly with J.A. and A.M.; J.A. and K.P.N. contributed to the analysis and interpretation; J.A., A.M., and K.N.P. directed and supervised the study. All authors reviewed the manuscript.

\section*{Additional Information}
\textbf{Competing Interests}: Authors declare no competing interests.

\end{document}